\documentstyle[12pt]{article}

\begin{document}

\begin{titlepage}

\begin{center}

\hfill  hep-th/9506024\\

\vskip .7in

{\bf A NOTE ON THE STRING ANALOG OF $N\!=\!2$ SUPER-SYMMETRIC
YANG-MILLS}

\vskip .3in

C\'esar G\'omez\footnote{Permanent address: Instituto de
Matem\'aticas y F\'isica Fundamental (C.S.I.C.), Serrano 123, 28006
Madrid, Spain.} and Esperanza L\'opez$^{\mbox{\footnotesize{1}}}$

\vskip .3in
{\em D\'epartement de Physique Th\'eorique, Universit\'e de
Gen\`eve,}

{\em  Gen\`eve 6, Switzerland}

\end{center}

\vskip .3in

\begin{center} {\bf Abstract}  \end{center}
A connection between the conifold locus of the type II string on the
$W\:P_{11226}^4$ Calabi-Yau manifold and the geometry of the quantum
moduli of $N\!=\!2$ $SU(2)$ super Yang-Mills is presented. This
relation is obtained from the anomalous behaviour of the $SU(2)$
super Yang-Mills special coordinates under $S$-duality transformation
in $Sl(2;Z)\!/\!\Gamma_2$.
\vskip 3.3in

\noindent June 1995

\end{titlepage}

{\it 1.} A string analog of $N\!=\!2$ supersymmetric Yang-Mills (SYM)
is obtained in reference \cite{KV} by means of a "rank three"
compactification of the heterotic string on $T^2\!\times\!K_3$. The
corresponding $d\!=\!4$ $N\!=\!2$ theory contains two vector
multiplets, one of them to be identified with the dilaton $S$, and an
extra $U(1)$ vector associated with the graviphoton. The basic claim
in \cite{KV} is the existence of a dual type II string
\cite{HT,W,S,HS} on the Calabi-Yau manifold $W\:P_{11226}^4\!=\!M$
\cite{C,Y} such that the quantum moduli of the heterotic string is
given by the moduli of the vector multiplets of the type II string on
$M$\footnote{It was also suggested in reference \cite{FC} that the
quantum moduli for a heterotic string of rank $r$ should look as the
complex moduli of a Calabi-Yau manifold with $h_{21}\!=\!r$.}, which
receives no quantum corrections. Accepting the previous framework, it
should be expected that the geometry of the quantum moduli of
$N\!=\!2$ $SU(2)$ SYM, solved in \cite{SW}, can be recovered from the
structure of the moduli of vector multiplets on $M$. In reference
\cite{KV} some evidence supporting this picture is presented.

The moduli of complex structures for the mirror of $M$ can be
parametrized by two complex variables $\phi$, $\psi$, defined by
\begin{equation}
P_{11226}= z_{1}^{12}+z_{2}^{12} + z_{3}^6 + z_{4}^6 +z_{5}^2 -
12 \psi z_1 z_2 z_3 z_4 z_5 -2 \phi z_{1}^6 z_{2}^6
\label{1}
\end{equation}
In terms of the coordinates ${\bar x}\!=\!\frac{-\phi}{864 \psi^6}$,
${\bar y}\!=\!\frac{1}{\phi^2}$ introduced in \cite{Y}, the locus of
conifold singularities for (\ref{1}) is given by
\begin{equation}
(1-{\bar x})^2 -{\bar x}^2 {\bar y} =0
\label{2}
\end{equation}
where $j$ is the Jacobi elliptic function.
For fixed ${\bar y}\!\neq\!0$, equation (\ref{2}) provides two
singular points ${\bar x}^{\pm}$. Identifying, at leading order in
the coupling, ${\bar y}\!=\!e^{-S}$, it was observed in \cite{KV}
that for $S\!\rightarrow\!\infty$
the conifold locus merges to the point ${\bar x}\!=\!1$. Using the
relation
found in \cite{C}
\begin{equation}
{\bar x}= \frac{1728}{j(\tau)}  \hspace{1cm} (j(i)=1728),
\label{3}
\end{equation}
the point ${\bar x}\!=\!1$ corresponds to $\tau\!=\!i$, which, for
$\tau$ the elliptic parameter of the internal torus used in the
heterotic compactification \cite{KV},  coincides precisely with the
$SU(2)$ point of enhanced gauge symmetry.
The picture emerging from this comment being that, for finite but
small ${\bar y}$, we should be able to derive the way in which the
classical singularity of $N\!=\!2$ $SU(2)$ SYM, where the $SU(2)$
symmetry is restored perturbatively, splits into the two singular
points of the quantum moduli \cite{SW}. It is important to keep in
mind that, in the case of $N\!=\!2$ $SU(2)$ SYM, the quantum moduli
is not the fundamental domain with respect to the full modular
(duality) $Sl(2;Z)$ group, while for the string analog, the elliptic
parameter $\tau$ in (\ref{3}) is, as a consequence of T-duality, in a
$Sl(2;Z)$ fundamental domain.

In this note we will approach the problem of connecting the quantum
moduli for $SU(2)$ SYM and the moduli of vector multiplets on $M$.
Our approach will consist in implementing the strong-weak coupling
duality transformations of $N\!=\!2$ SYM as elements in $Sp(6;Z)$.
This will imply a deep connection between the dilaton and the
geometry of the quantum moduli of SYM, from which we will try to
derive the conifold locus of the Calabi-Yau manifold (\ref{1}).

\vspace{1cm}

{\it 2.} In \cite{U} we have shown that the obstruction to
strong-weak coupling duality invariance in $N\!=\!2$ SYM, comes from
the fact that\footnote{From now on we will use the notation of
reference \cite{SW}.}
\begin{equation}
a(\gamma(u)) \neq a_D (u)
\label{5}
\end{equation}
where the transformation $\gamma$ maps a strong-coupling singular
point into the asymptotically free point at infinity. However the
effective gauge coupling parameter $\tau
\!=\!i\frac{4\pi}{g_{e\!f\!f}^2}+\frac{\theta_{e\!f\!f}}{2\pi}$
behaves in a dual-like way
\begin{equation}
\tau(\gamma(u)) = -\frac{1}{\tau(u)}
\end{equation}
For the curve \cite{SW} which parametrized the quantum moduli of
$N\!=\!2$ $SU(2)$ SYM
\begin{equation}
y^2= (x+\Lambda^2)(x-\Lambda^2)(x-u)
\label{4}
\end{equation}
with $\Lambda$ the dynamically generated scale, the transformation
$\gamma$ is given by
\begin{equation}
\gamma(u) = \Lambda^2 \: \frac{u+3\Lambda^2}{u-\Lambda^2}
\label{10}
\end{equation}
which maps the singular point $u\!=\!\Lambda^2$ into $\infty$.
This transformation is an element of
$\Gamma_W\!=\!Sl(2;Z)\!/\!\Gamma_2$, which consists in
transformations of the coordinates $(x,y)$ of the curve that can be
compensated by a change in the moduli parameter $u$.

After coupling to gravity, i.e. in non-rigid special geometry, we can
improve (\ref{5}) to \cite{U}
\begin{equation}
a^{gr}(\gamma(u)) = {\tilde f}_{\gamma, \Lambda} (u)\: a_{D}^{gr} (u)
\label{7}
\end{equation}
if at the same time we perform the transformation
\begin{equation}
a_0 (u) \rightarrow {\tilde f}_{\gamma, \Lambda} (u) \: a_0 (u)
\label{9}
\end{equation}
with
\begin{equation}
{\tilde f}_{\gamma, \Lambda} (u)= \frac{\sqrt 2}{4\pi} \left(
\frac{2\Lambda^2}{\Lambda^2-u} \right)^{3/2}
\label{13}
\end{equation}
Let us explain briefly this result.
In the context of rigid special geometry, the 1-form $\lambda(x,u)$,
whose periods define the holomorphic sections $(a,a_D)$, satisfies
\begin{equation}
\frac{d\lambda}{du} \sim \lambda_1
\label{6}
\end{equation}
with the proportionality factor determined by imposing the physically
correct asymptotic behaviour at the singularities, and where
$\lambda_1(x,u)$ is the everywhere non-zero holomorphic 1-form of the
curve (\ref{4}).
The 1-form $\lambda^{\gamma}(x,u)$ which defines $(a(\gamma(u),
a_D(\gamma(u))$, also verifies equation (\ref{6}) with
\begin{equation}
\frac{d\lambda^{\gamma}}{du} = {\tilde f}_{\gamma, \Lambda}\lambda_1
\label{8}
\end{equation}
In order to pass from (\ref{5}), where the relation between
$a(\gamma(u))$ and $a_D(u)$ is not a symplectic transformation, to
(\ref{7}) which, up to the holomorphic function ${\tilde f}_{\gamma,
\Lambda}$, is symplectic, we can replace the derivative in (\ref{8})
by a covariant derivative relative to the graviphoton (Hodge)
$U(1)$-connection \cite{U}. From this, equation (\ref{9}) immediately
follows. However, the gauge transformation ${\tilde f}_{\gamma,
\Lambda}$ is singular and generates a topological obstruction to
implement strong-weak coupling duality. Thus, the more natural way to
interpret (\ref{7}) and (\ref{9}) would be to add an extra dilaton
field $S$.

Namely, for the "rank three" heterotic string compactification on
$T^2\!\times \!K_3$ we can parametrize the two vector multiplets by
the special coordinates $(X^0,X^1,X^2)$, with
$S\!=\!-i\!X^1\!/\!X^0)$.
As it was shown in \cite{K,NF}, in order to have all the couplings
proportional to $S$, it is necessary to do the following symplectic
change of variables
\begin{eqnarray}
{\hat X}^{I \neq 1} = X^I &,& {\hat F}_{I \neq 1} = F_I \nonumber \\
{\hat X}^1 = F_1 &,& {\hat F}_1 = -X^1 = - i S X^0
\end{eqnarray}
In these new variables, we can interpret transformation (\ref{9}) as
\begin{equation}
{\hat X}^0 \rightarrow  i S {\hat X}^0 = - {\hat F}_1
\label{12}
\end{equation}
Performing twice the transformation $\gamma$ given in (\ref{10}), we
get the identity (up to a sign ambiguity). This implies that under
$\gamma$ satisfying (\ref{9}), the dilaton
changes as
\begin{equation}
S\rightarrow S'= \frac{1}{S}
\end{equation}
At the level of the special coordinates, this equation implies that
${\hat F}_1 \! \rightarrow \!X^0$, which very likely, together with
(\ref{12}), will
be part of a symplectic transformation in $Sp(6;Z)$.

Interpreting now equation (\ref{9}) as equation (\ref{12}) boils down
to the following formal identification
\begin{equation}
S \sim {\tilde f}_{\gamma , \Lambda} (u)
\label{14}
\end{equation}

\vspace{1cm}

{\it 3.} Our next task will be to give a natural physical meaning to
relation
(\ref{14}). Before doing that it would be worth to come back to our
original problem. Once we have added the dilaton field $S$, we can
parametrize the moduli of the two vector multiplets by $u$, the
coordinate of the $N\!=\!2$ SYM quantum moduli, and $S$. Naively this
looks like a one parameter family of $u$-planes labeled by $S$. The
natural physical picture, according with the identification of the
conifold locus for $S\!\rightarrow\!\infty$ with the classical
singularity of enhanced $SU(2)$ symmetry, will be to have, for
$S\!\rightarrow\!\infty$, a classical moduli with one singularity at
$u\!=\!0$ and, for finite $S$, the quantum moduli with two
singularities at $u\!=\!\pm\!\Lambda^2$, where $\Lambda$ will now
depend on the expectation value of the dilaton. However, the previous
picture is too crude since it simple fibers the $N\!=\!2$ SYM quantum
moduli on the dilaton line. This reduces the effect of the dilaton to
simply changing the scale $\Lambda$, and in this way, the location of
the two singularities. Our approach will be to parametrize the moduli
space of the two vector multiplets by $(u,S)$ variables, and to try
to recover the conifold locus (\ref{2}) directly from relation
(\ref{13}), which we have derived from the geometry of the quantum
moduli of reference \cite{SW}, by interpreting the transformations
(\ref{7}) and (\ref{9}) as the symplectic transformation (\ref{12})
of the heterotic string.

In order to use (\ref{14}) for defining a locus in the $(u,S)$
variables, we need first to obtain some explicit relation between the
dynamical scale $\Lambda$ and the value of the dilaton field $S$.
Using that the unrenormalized gauge coupling constant $g_0$ of the
effective field theory derived from an heterotic string is
proportional to the dilaton expectation value and that changes in
$g_0$ can be absorbed in changing the scale $\Lambda$, we will
impose, based on the renormalization group equation for $N\!=\!2$
$SU(2)$ SYM, the relation
\begin{equation}
\Lambda^2 = e^{-S/2}
\end{equation}
Combining this with (\ref{13}), equation (\ref{14}) becomes
\begin{equation}
\left( \frac{4 \pi}{\sqrt 2}  S \right)^{2/3} = \frac{2
e^{-S/2}}{e^{-S/2}-u}
\label{15}
\end{equation}
which can now be used to define a locus $S(u)$ in the $(u,S)$ space.
We claim that the so-defined locus in $(u,S)$ variables should
coincide with the conifold locus (\ref{2}).

{}From (\ref{15}) we get
\begin{equation}
e^{-S/2} = \frac{u}{1- 2\left(\frac{\sqrt 2}{4 \pi S} \right)^{2/3}}
\label{16}
\end{equation}
For $S\!\rightarrow\! \infty$ we can use, at leading order, the
identification ${\bar y}\!=\!e^{-S}$ \cite{KV}. Solving the conifold
locus (\ref{2}), we obtain
\begin{equation}
{\bar x}^{\pm} = \frac{1}{1 \pm \sqrt{\bar y}}
\end{equation}
which, using (\ref{16}), can be writen as
\begin{equation}
{\bar x}^{\pm} =  \frac{1- 2\left(\frac{\sqrt 2}{4 \pi S(u)}
\right)^{2/3}}{1\pm u  - 2\left(\frac{\sqrt 2}{4 \pi S(u)}
\right)^{2/3}}
\end{equation}
Notice that this equation is equivalent to postulate that the locus
(\ref{16})
is equivalent to the conifold locus (\ref{2}).
However, motivated by this equation we can propose a map from the
$(u,S)$ variables into the $({\bar x},{\bar y})$ variables
parametrizing the complex moduli of (\ref{1})
\begin{equation}
{\bar x} (u,S) =  \frac{1- 2\left(\frac{\sqrt 2}{4 \pi S}
\right)^{2/3}}{1 + u  - 2\left(\frac{\sqrt 2}{4 \pi S} \right)^{2/3}}
\label{17}
\end{equation}

We need now to check the consistency of (\ref{17}). First we observe
that it reflects correctly the global $Z_2$ symmetry of the $N\!=\!2$
$SU(2)$ SYM gauge theory. The action of this global symmetry in the
SYM quantum moduli is given by $u\!\rightarrow\!-\!u$, therefore
mapping the two finite singular points between themselves. We observe
that the change $u\!\rightarrow\!-\!u$ in (\ref{17}) also maps ${\bar
x}^+$ into ${\bar x}^-$ and vice-versa.

Second, in the string weak-coupling limit $S\!\rightarrow\!\infty$,
we get
\begin{equation}
{\bar x}= \frac{1}{u+1} + {\cal O}( S^{-2/3})
\label{18}
\end{equation}
which implies that the asymptotically free regime of the $N\!=\!2$
SYM theory, which corresponds to $u\!\rightarrow\!\infty$, is
associated with ${\bar x}\!=\!0$\footnote{In reference \cite{KV} is
is also suggested that ${\bar x}\!=\!0$ can be interpreted as a weak
coupling region related to the degeneration of the compactification
torus.}.
Using now the relation ${\bar x}\!=\!\frac{-\phi}{864 \psi^6}$, we
obtain that $\frac{1}{u}\!=\!\frac{-\phi}{864 \psi^6}$ in a
neighborhood of the asymptotically free regime
($u\!\rightarrow\!\infty$). In order to compare the instanton
expansion of $SU(2)$ SYM with the expansion in the $(\phi,\psi)$
variables of the periods of the holomorphic top form for the
Calabi-Yau (\ref{1}) \cite{C}, we will consider $\Lambda$ and
$e^{-S}$ as independent parameters defining the following doble
limit\footnote{In reference \cite{KV} it has been already proposed to
use a suitable "doble scaling" limit in order to recover the
$N\!=\!2$ $SU(2)$ SYM quantum moduli from the moduli space of complex
deformations of the $W\:P_{11226}^4$ Calabi-Yau.}
\begin{equation}
lim_{\Lambda, S \rightarrow \infty}
\Lambda^4 e^{-S} =1
\end{equation}
The instanton expansion of $SU(2)$ SYM goes like $\left(
\frac{1}{a(u)} \right)^{4k}$, $k\!>\!0$. Using that $a\!=\!\sqrt{2u}$
when $u\!\rightarrow\!\infty$ \cite{SW}, in the $u$ variable, we get
$\left( \frac{1}{u} \right)^{2k}$. On the other hand, the leading
term as $\phi\!\rightarrow\!\infty$ in the expansion of the periods
of the top form behaves as $\left( \frac{\phi}{\psi^6} \right)^{k}$.
The presence in one case of powers of $2k$ and in the other of powers
of $k$ can be explained by the fact that,
for the Calabi-Yau case, the two singular points of the locus
(\ref{2}) are related by the transformation
$\phi\!\rightarrow\!-\!\phi$ which is not a symmetry, while the $Z_2$
transformation $u\!\rightarrow\!-\!u$, that corresponds by (\ref{16})
to $\phi\!\rightarrow\!-\!\phi$, is a global symmetry of $N\!=\!2$
$SU(2)$ SYM. Therefore, in order to reproduce the SYM quantum moduli
inside the Calabi-Yau moduli space, we should mod by the
transformation
$\phi\!\rightarrow\!-\!\phi$, or equivalently, consider only even
powers of
$\frac{\phi}{\psi^6}$.

Summarizing, we have shown that by identifying the conifold locus
with equation (\ref{14}) we get, at leading order in $S$, a
consistent physical picture. The important thing for us is that
equation (\ref{14}) was obtained by promoting, introducing a dilaton
field, the anomalous strong-weak coupling duality transformations of
the $N\!=\!2$ SYM field theory, to good symplectic transformations.
In other words the quantum moduli in reference \cite{SW}
describes more than the complex structure of the curve (\ref{4}),
something which is related to the existence of a dynamically
generated scale $\Lambda$. When we include the dilaton and work in
the string framework, what was geometrical (no moduli) information
for $SU(2)$ SYM becomes now pure moduli information of the Calabi-Yau
manifold (\ref{1}).

\vspace{1cm}

This work was partially supported by
european community grant ERBCHRXCT920069, by PB 92-1092 and by OFES
contract number 93.0083. The work of E.L. is supported by a
M.E.C. fellowship AP9134090983.

\end{document}